# Quantification of heparin in complex matrices (including urine) using a mix-and-read fluorescence assay

Ulrich Warttinger[1], Roland Krämer[1]

Correspondence to:
Roland Krämer, phone 0049 6221 548438, fax 0049 6221 548599
E-mail: kraemer@aci.uni-heidelberg.de

1 Heidelberg University, Inorganic Chemistry Institute, Im Neuenheimer Feld 270, 69120 Heidelberg, Germany.

**Abstract**

Heparin is an important anticoagulant drug, about one billion doses are produced annually. It is a polydisperse sulfated polysaccharide, and the inherent heterogeneity makes the analysis of heparin difficult. The global crisis resulting from adulterated heparin in 2008 has drawn renewed attention to the challenges that are associated with the quality control and characterization of this complex biological medicine from natural sources. The present study addresses the need for simple and user-friendly analytical methods for the fast and accurate quantification of heparin in complex matrices. Direct quantification of heparin in the low microgram per mL range was accomplished using a specific commercially available assay based on the fluorescent molecular probe Heparin Red, simply by mixing the heparin containing sample and a reagent solution in a 96-well microplate followed by fluorescence readout. A screening of typical impurities in raw heparin (selected other glycosaminoglycans, residual nucleic acids and proteins), related to the extraction from animal tissues, as well as of components of the urine matrix (inorganic salts, amino acids, trace proteins) revealed that these compounds even in large excess have no or very little effect on the accuracy of heparin determination. Heparin spike detection in urine, a biological multicomponent matrix, also showed good accuracy. We envision applications of this mix-and-read assay in the process and quality control in heparin manufacturing, but also in pharmacokinetic studies as a convenient tool for measuring of the urinary excretion of heparins.

## Introduction

Heparins are widely used clinical anticoagulants [1, 2], about one billion doses are produced each year.[3] Heparin is a polydisperse mixture of sulfated linear polysaccharides that consist of disaccharide repeating units (scheme 1, left) and have a high negative charge density of about -1.7 per monosaccharide. Unfractionated heparin (UFH, mean molecular weight between 13000 and 15000), is clinically applied since the 1930s. In recent decades, there has been a trend towards use of low molecular weight heparins (LMWH, mean molecular weight between 4000 and 7000). The latter are prepared from UFH by partial depolymerisation and display more favourable pharmacokinetics and lower side effects. In the following, the term "heparin" refers to unfractionated heparin.

Heparin is prepared by extraction from the tissues of slaughterhouse animals, mainly porcine intestine. The process involves protein hydrolysis at alkaline pH in the presence of proteases and enrichment of the heparin like GAGs by anion exchange resin.[4] The raw or crude heparin obtained at this stage is typically a mixture of many components – a rough estimate is 50% heparin and 50% other compounds [5], mostly other glycosaminoglycans of lower sulfation degree (such as dermatan sulfate and chondroitin sulfate) but also residual biomaterials including proteins, peptides, nucleic acids, and lipids. Preparation of raw heparin is typically performed outside of GMP (Good Manufacturing Practices). There is a significant process-dependent variation in the composition of raw heparin. The crude material is then converted under GMP conditions to pharmaceutical grade heparin in a further multistep purification process, involving precipitation, ion exchange chromatography and chemical treatment. Monitoring the raw heparin preparation and purification process is challenging, and new, effective analytical methods for fast in-process quantification of heparin have the potential to improve production efficiency and reduce waste.[6]

In 2008, the economically motivated adulteration of raw heparin with oversulfated chondroitin sulfate, a substance that mimics the anticoagulant activity of heparin but is much cheaper to produce, was associated with about 200 deaths and hundreds of adverse advents worldwide. This "heparin crisis" has drawn renewed attention to the challenges that are associated with the characterization, quality control and standardisation of complex biological medicines from natural sources.[7] The tragic event resulted in the introduction of advanced analytical screening of pharmaceutical heparin, including NMR spectroscopy, capillary electrophoresis , and HPLC. To ensure continued safety of heparin, the US Food and drug administration (FDA) recommended continued development of additional methods for the quality control of heparin.

In clinical settings, heparin blood levels are monitored by the antithrombin-mediated effect on the clotting cascade or individual coagulation factors. These assays can, however, not be readily transferred to other complex matrices. Rather, heparin quantification in urine, for example, is achieved by challenging and time consuming protocols that either involve radiolabeling [8, 9] or require the isolation of heparin from urine and staining/colorimetric procedures, eventually including electrophoresis [10, 11]. Depending on the methodology, urinary recovery of heparin between 1% and 39 weight % of injected doses has been reported. In one study [10], roughly one half of the excreted was characterized as unchanged heparin, and the other half as a partially desulfated metabolite with a charge density of about -1.5 per monosaccharide. There is a renewed recent interest in non-invasive (in particular oral) delivery options for heparin [12, 13], and quantifying urinary excretion is part of the pharmacokinetic analysis.[14] Oral uptake of a 1000 U/kg body weight heparin dose resulted in urinary concentrations between 0,2 and 4 µg/mL. [14]

There is a need for simple and user-friendly analytical methods for fast and accurate quantification of heparin in complex matrices. Such methods would be of great value for process control in crude heparin preparation and purification, assessing the comparibility of raw heparin batches in terms of heparin content, and for heparin quantification in complex biological matrices other than blood samples. A key requirement on such an assay is elimination of matrix effects by suppressing the potential interference of a multitude of abundant matrix components. Cationic dyes, such as Alcian Blue and Azure A have historically been used to detect heparin colorimetrically, but it is well known that many other substances interfere with the non-covalent binding that underpins these assays.[15]

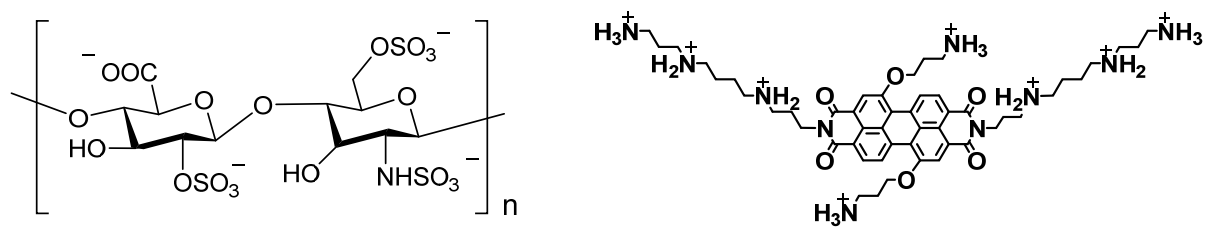

**Scheme 1**. Major repeating disaccharide unit of heparin (left). Structure of the polycationic fluorescent probe Heparin Red (left).

We describe here the application to heparin detection in various matrices of the fluorescent probe "Heparin Red", initially described as an experimental probe for heparin detection [16] and meanwhile developed further into commercially available assays. Heparin Red is a polyamine derivative of a red-emissive perylene diimide fluorophore (scheme 1, right). It forms a supramolecular complex with the target, with aggregation of the probe molecules at the heparin template and contact quenching of fluorescence (scheme 2). The strong binding of the polycationic probe to polyanionic heparin appears to be controlled by both electrostatic and aromatic pi-stacking interactions [17].

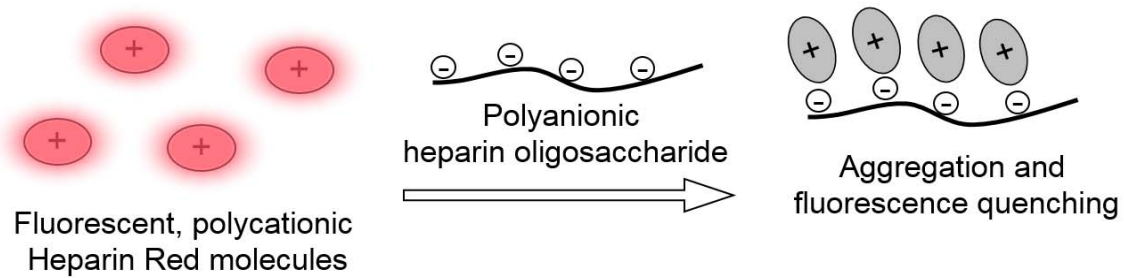

**Scheme 2.** Schematic representation of fluorescence quenching of the molecular probe Heparin Red in the presence of polyanionic heparin.

We have recently applied the commercial Heparin Red Kit, optimized for blood plasma matrix, to the direct quantification of heparins (including unfractionated heparin, low molecular weight heparin and non-anticoagulant heparins) and fucoidan (a polysulfated polysaccharide from brown algae) in human plasma.[18, 19] The Kit is of particulat value for non-anticoagulant heparins that can not be analyzed by standard coagulation assays and is emerging as a tool for pharmacokinetic analysis of this promising class of drug candidates.[20] While the Heparin Red Kit provides convincing results in the highly complex plasma matrix, it is prone to interference by nucleic acids (ususally not present in blood plamsa at signicant concentrations) and does not give satisfactory results for heparin detection in urine.

In this contribution, we describe for the first time the application of Heparin Red Ultra, a more recently released commercial assay, to heparin quantification in various matrices including urine. Heparin Red Ultra has a different reagent formulation and is provided as a ready-to-use solution that simply needs to be mixed with the heparin containing sample. The study includes a systematic screening of the effect of individual components of raw heparin preparations and of the urine matrix on the accuracy of the assay.

## Materials and Methods

### Instrumentation

*Fluorescence measurements*

Fluorescence was measured with a microplate reader Biotek Synergy Mx (Biotek Instruments, Winooski, VT, USA), excitation at 570 nm, emission recorded at 605 nm, spectral band width 17 nm, gain 110, read height of 8 mm.

*Microplates*

For fluorescence measurements 96 well microplates, polystyrene, Item No 655076, were purchased from Greiner Bio-One GmbH, Frickenhausen.

*Pipettes*

Transferpette® 0,5-10µl, Transferpette®-8 20-200µl and Transferpette®-12 20-200µl, purchased from Brand GmbH, Wertheim. Rainin Pipettes 100-1000µl, 20-200µl, and 2-20µl purchased from Mettler Toledo, OH, USA.

### Reagents

Heparin Red® Ultra was a gift from Redprobes UG, Münster, Germany. Product No HRU001, Lot 003.

Aqueous solutions were prepared with HPLC grade water purchased from VWR, product No 23595.328. Unfractionated heparin sodium salt from porcine intestine mucosa was purchased from Sigma-Aldrich GmbH, Steinheim, product number H5515, Lot SLBK0235V. Heparin stock solutions were stored at 4-8°C. Inorganic salts were >99% pure. Chondroitin sulfate A sodium salt from bovine trachea, product number C9819, Lot SLBQ0017V, 85% pure according to 13C NMR spectrum (Certificate of analysis); Dermatan sulfate sodium salt from porcine intestinal mucosa, 100% pure according to 13C NMR spectrum (Certificate of analysis), product number C3788, Lot SLBM9912V; DNA sodium salt from salmon testes, product number D1626, Lot SLBF9870V; RNA from yeast, product number *10109223001 Roche*, Lot 13833221; Glycine, product number G5417, Lot CDBC1274V; Immunogloblin G (IgG) from human serum, product number I4506, Lot SLBK8678V, 97% pure according to SDS electrophoresis (Certificate of analysis); "Protease" isolated from *Streptomyces griseus*, lyophilized powder, caseinolytic activity >3.5 units/mg, mixture of several proteinolytic activities (serin type proteases including trypsin, zinc endopeptidase, zinc aminopeptidase, zinc carboxypeptidase), product number P5147, Lot SLBQ0332V; Human serum albumin, product number A1653, Lot SLBD1834V, 99% pure according to agarose electrophoresis (Certificate of analysis); and Tris(2-carboxyethyl)phosphine (TCEP), product number 646547, Lot MKBZ1912V, as 0.5 M solution, pH 7.0 adjusted with ammonium hydroxide, were all purchased from Sigma Aldrich GmbH, Steinheim, Germany.

*Urine*

Urine samples were collected from two voluntary healthy donors, stored at -20°C and thawed before use. Heparin spiked urine samples were either used directly after spiking, or stored at -20°C and thawed prior to use.
  .

**Assays**

*Heparin Red® Ultra*

For determination of heparin concentrations in aqueous or urine samples, the protocol of the provider for a 96-well microplate assay was followed. In brief, 5 µL of the heparin spiked sample was pipetted into a microplate well. Then, 180 µL of Heparin Red Ultra solution was added to the samples as simultaneously as possible. For samples numbers > 10, a 12-channel pipette was used for addititon of Heparin Red Ultra solution. The microplate was immediately introduced in the fluorescence reader and mixing was performed using the plate shaking function of the microplate reader (setting "high", 1 minute). Immediately after mixing, fluorescence was recorded within 1 minute. Detections in water and urine were performed as either duplicates or triplicates.

*HemoCue® albumin 201 system*

The Hemocue albumin 201 system is a portable, lab-accurate system for quantification of human serum albumin in urine. For determination of albumin concentrations in µg/mL, the protocol of the provider was followed.

## Results and discussion

### Detection of heparin in complex matrices

The Heparin Red Ultra assay was first applied to the detection of heparin at different concentrations in aqueous samples. Figure 1 shows the decrease of fluorescence signal with increasing heparin concentration. Response shows good linearity in the range 0-7 µg/mL.

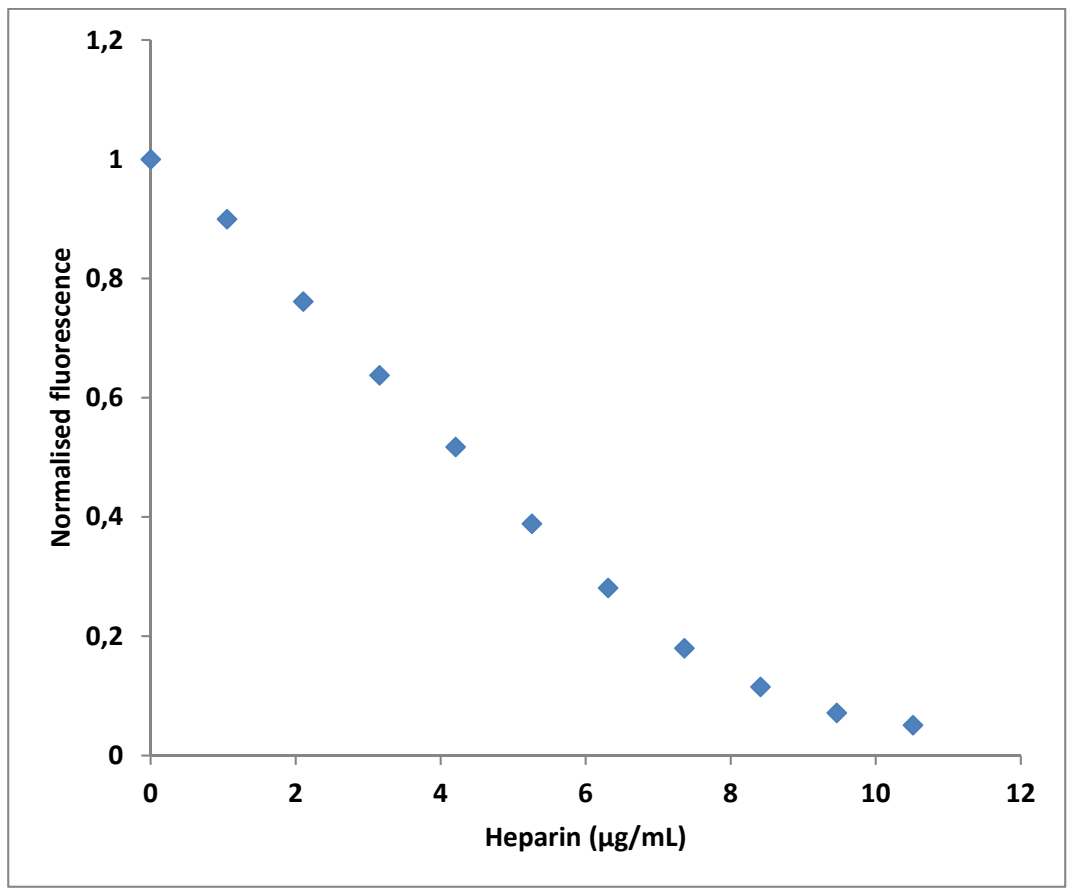

**Figure 1.** Normalized response curve of the Heparin Red Ultra fluorescence assay to aqueous heparin samples in the concentration range 0-10 µg/mL. Manually performed microplate assay. Excitation at 570 nm, fluorescence emission at 605 nm. Averages of duplicate determinations; CV for the linear range 0 to ≈7 µg/mL (averaged over all concentrations): 1,7 %.

Analytical precision was determined for selected heparin concentrations, CVs relating to fluorescence internsity were 0.3% for c(heparin) = 0 µg/mL (n=8) and 2.9 % for c(heparin) = 4,38 µg/mL (n=8). At the latter concentration, an about 50% decrease of fluorescence intensity is observed.

It should be noted at this point that, as described previously for the Heparin Red Kit [19], the sensitivity of the assay is tunable by modification of the protocol. Increasing the sample volume from 5 µL (see "Materials and Methods", "Assays") to 10 µL enhances the sensitivity about twofold, i.e. only 0.4 µg/mL heparin trigger a 10% fluorescence reduction (data not shown). Also, the low molecular weight heparin enoxaparin is in 10 µL samples detected with good sensitivity (about 50% fluorescence reduction by 5 µg/mL, data not shown). The gain in sensitivity might, however, be at the cost of accuracy of the assay in the presence of potentially interfering compounds (figure 2).

To elicit the performance of the assay for heparin quantification in diverse matrices, we have repeated the assay at two heparin concentrations, 0 µg/mL (absence of heparin) and 4,38 µg/mL (about 50% reduction of fluorescence signal) in the presence of individual substances that are potential components of crude heparin preparations or of the urine matrix. These substances include inorganic salts, other glycosaminoglycans, proteins, amino acids and nucleic acids.

Results are shown in figure 2. The control sample (leftmost in the diagram) refers to water and 4,38 µg/mL heparin in water, respectively. Fluorescence in the presence of most of the added substances at specified concentrations is within a narrow range, as indicated by the grey horizontal bars, of ± 2.6 % (heparin free samples) or ± 3.1 % (heparin spiked samples) from control, means no or very little interference. We conclude that the Heparin Red Ultra assay suppresses a wide variety of potential interferences in complex matrices.

In detail, the assay tolerates 150 mM NaCl, 75 mM KCl, and inorganic salts of di/trivalent ions (including $Ca^{2+}$, $Mg^{2+}$, sulfate and phosphate) in the 10-50 mg/mL range, nucleic acids (DNA, RNA) at 25-50 µg/mL, the glycosaminoglycans chondroitin sulfate an dermatan sulfate at 25 µg/mL, the amino acid glycine at 20 mM, immunoglubulin G (a major plasma protein, also present in urine in trace amounts) at 30 µg/mL, and a commercial protease preparation at 2 mg/mL (proteases are extensively used in the manufacture of crude heparin). Only human serum albumin (HSA), the most abundant plasma protein and potential trace protein in urine, strongly masks fluoresecence response to heparin in concentrations as low as 30 µg/mL.

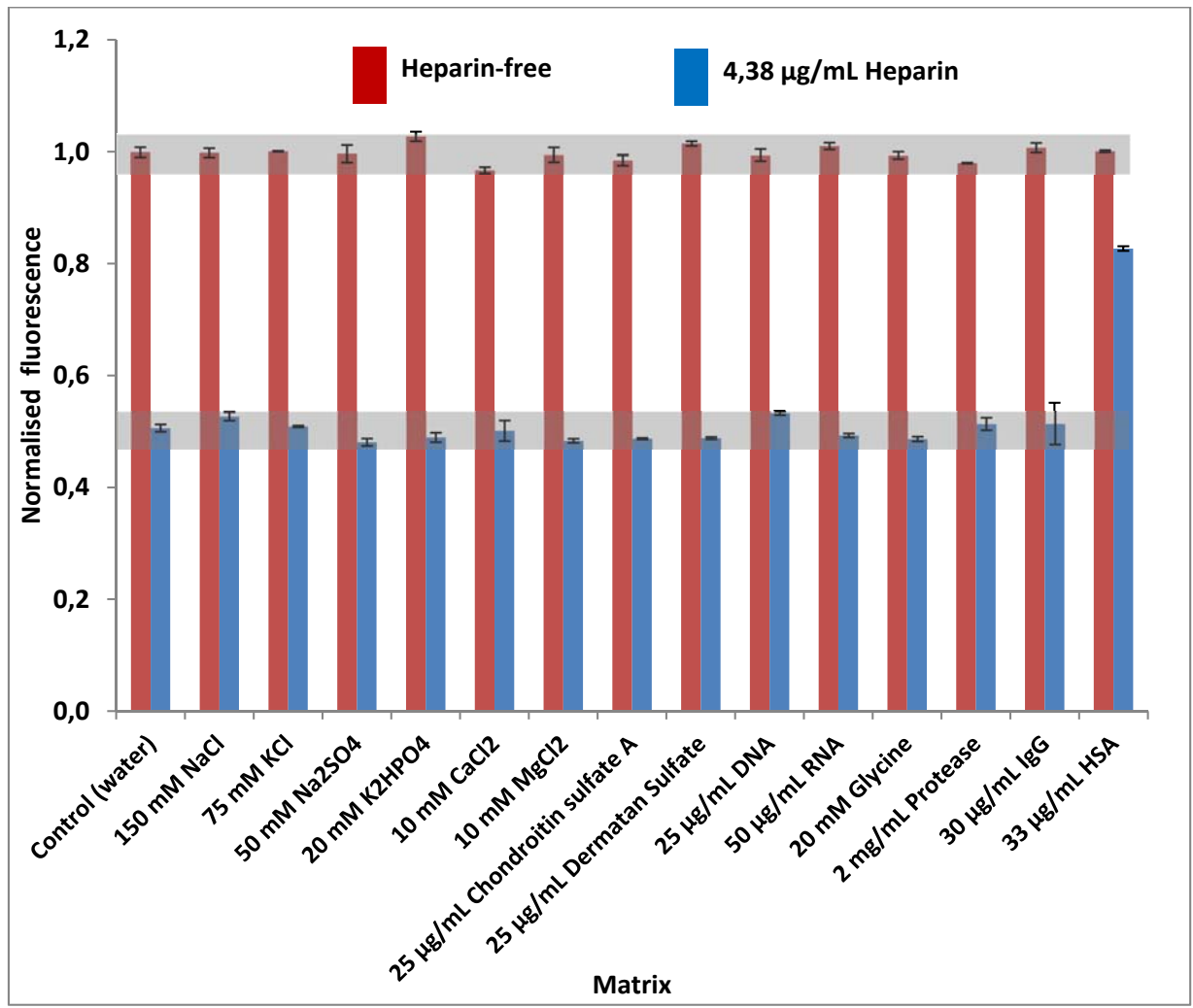

**Figure 2.** Fluorescence response of the Heparin Red Ultra assay to a heparin-free aqueous sample (red bars) and a sample containing 4,38 µg/mL heparin (blue bars), in the presence various substances at specified concentrations. Control (leftmost bars) refers to water or 4,38 µg/mL heparin in water, respectively. Averages of duplicate determinations with indicated error bars. IgG: Immunoglobuline G. HSA: Human serum albumin.

Interference by human serum albumine was analyzed in more detail. The masking effect increases with increasing concentrations of the protein (figure 3). The interference by 20 µg/mL HSA is completely eliminated by protease pretreatment of the heparin containing sample (commercial protease mix 1 mg/mL, 1 h at ambient temperature; data not shown), suggesting that only the intact HSA is masking heparin detection.

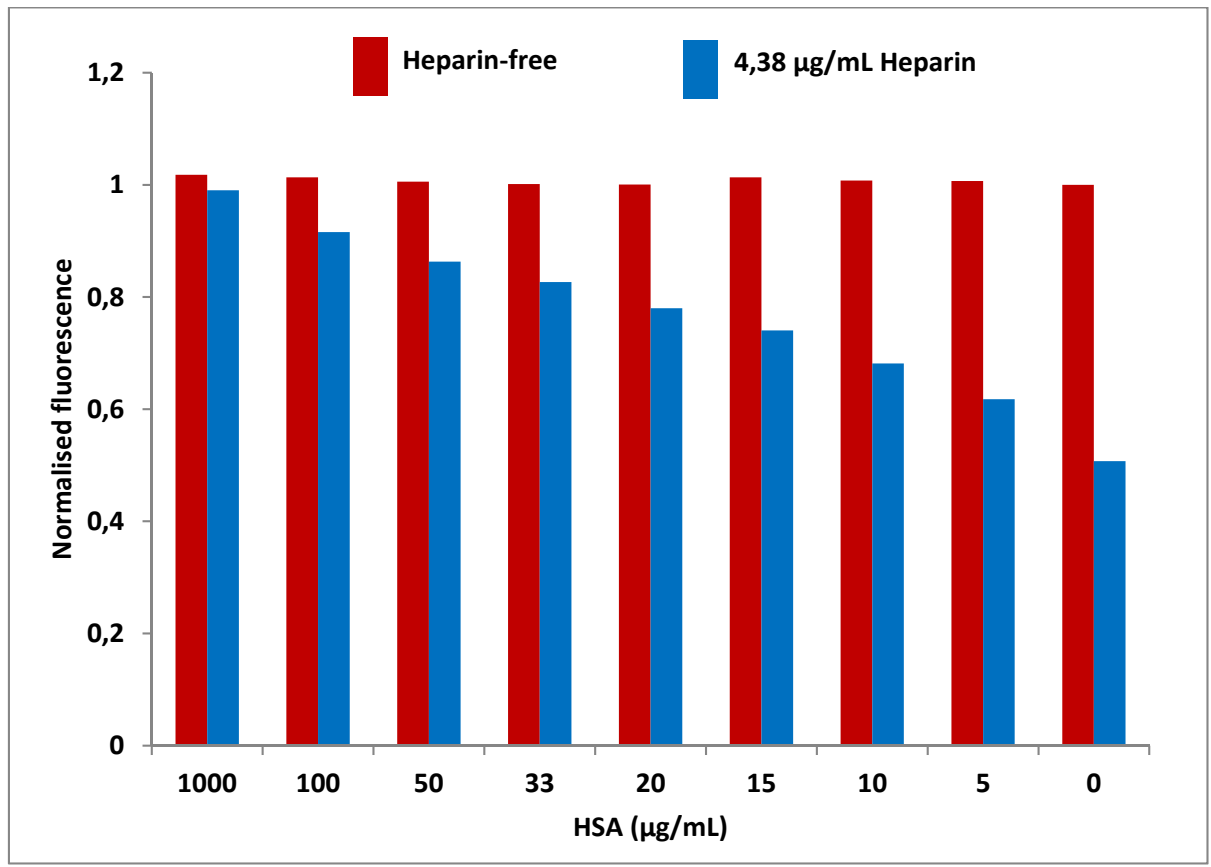

**Figure 3.** Fluorescence response of the Heparin Red Ultra assay to a heparin-free aqueous sample (red bars) and a sample containing 4,38 µg/mL heparin (blue bars), in the presence of human serum albumin (HSA) in varying concentrations.. Control (rightmost bars, or c(HSA) = 0 µg/mL) refers to water or 4,38 µg/mL heparin in water, respectively. Averages of duplicate determinations; CV averaged over all concentrations: 0.5 %.

Our hypothesis about the mechanism of HSA interference is based on protonation of carboxylate side chains (of Glu and Asp) in HSA in the acidic reaction medium provided by the Heparin Red Ultra assay. While the detailed composition of the reagent solution is not given by the provider, we have measured a pH value of 3.0 in a 1:1 mixture of Heparin Red Ultra solution and water with a standard glass electrode. At such a low pH value, the carboxylate side chains of Glu and Asp are expected to become protonated. Along with this, the 585 amino acid protein HSA would be converted from a charge-balanced zwitterionic species into polycationic polymer having an overall charge of +99 (due to the cationic residues of Lys, Arg and His).[21] This polycationic polymer may compete with the Heparin Red molecule for heparin binding and thus suppress the fluorescence response. It is not obvious, however, why only HSA should exhibit competitive heparin binding at low concentration but not immunoglobulin G or the protease (figure 2).

We therefore speculated that a specific structural feature of HSA triggers its strong masking effect. An unusual feature of HSA is the presence of 17 intramolecular disulfide bonds between Cys thiol residues.[21] This creates a number of intramolecular loops within the polypeptide chain in which the cationic amino acid residues are preorganized, possibly leading to a strong binding of the polyanionic heparin molecules. To support this idea, we incubated the HSA containing samples with the disulfide cleaving agent tris(2-carboxyethyl)phosphine (TCEP), 25 mM at ambient temperature for 15 minutes.[22] While TCEP itself does not effect the response of Heparin Red Ultra to heparin, it is able to eliminate the masking effect of HSA (figure 4). Since TCEP is commercially available in form of a "ready-to-use" solution, we found at the same time a very convenient sample pre-treatment method to overcome HSA interference in heparin detection by Heparin Red Ultra.

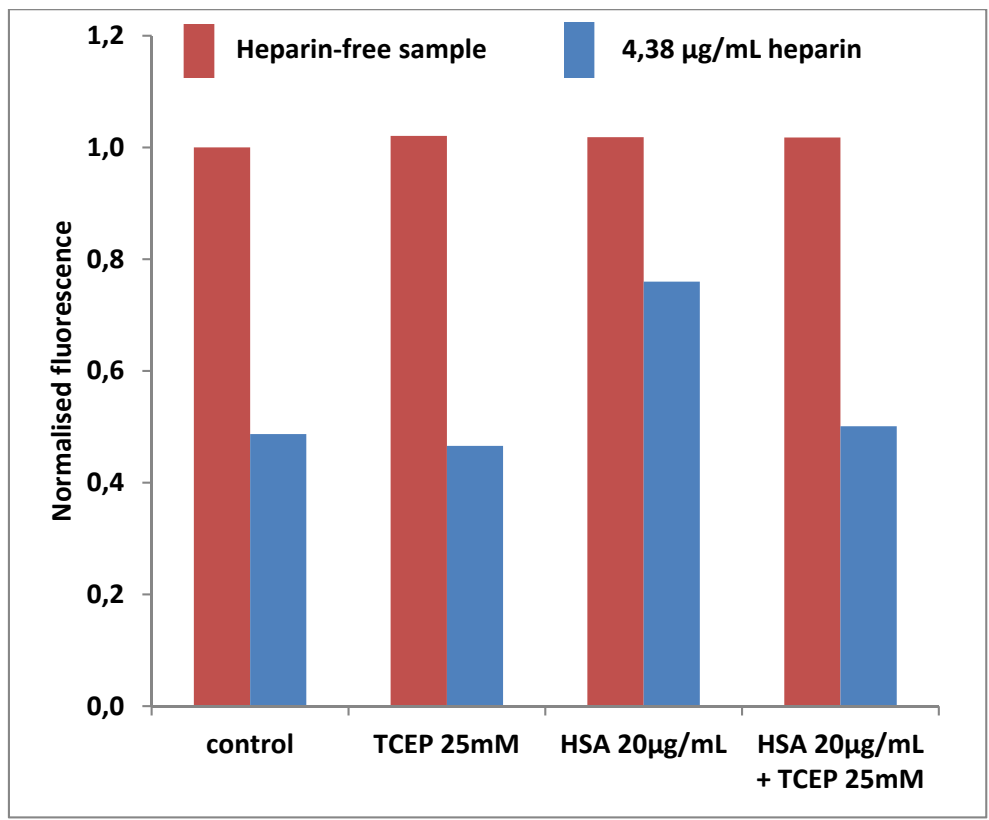

**Figure 4.** Fluorescence response of the Heparin Red Ultra assay to a heparin-free aqueous sample (red bars) and a sample containing 4,38 µg/mL heparin (blue bars), in matrices containing 20 µg/mL HSA and/or 25 mM TCEP (incubation time: 15 minutes). Control (leftmost bars) refers to water or 4,38 µg/mL heparin in water, respectively. Averages of duplicate determinations; CV averaged over all concentrations: 1.4 %.

**Detection of heparin in urine**

Urine is a complex biological matrix. It is a multicomponent aqueous mixture of highly variable concentrations of inorganic salts, organic compounds and trace amounts of bioacromolecules including proteins and gylcosaminoglycans. Table 1 lists the normal population reference ranges of selected urine components, in comparison with their non-interfering levels in the Heparin Red Ultra assay as shown in figure 2. The focus is on charged urine components that may pontentially interfere with the electroststic interaction between Heparin Red and heparin (compare scheme 2).

| Compound | Reference range for urine [23] | Non-interfering level in fluorescence assay |
|---|---|---|
| $Na^+$ | 27-167 mM | 150 mM |
| $Mg^{2+}$ | 2.5-8.5 mM | 10 mM |
| $Ca^{2+}$ | 2.5-7.5 mM | 10 mM |
| $K^+$ | 23-67 mM | 75 mM |
| $Cl^-$ | 67-167 mM | 150 mM |
| $SO_4^{2-}$ | 5-31 mM | 50 mM |
| $HPO_4^{2-}$ | 9-28 mM | 20 mM |
| CS | 0.5-10 µg/mL [24] | 25 µg/mL |
| DS | n.d. [24] | 25 µg/mL |
| DNA | < 1 µg/mL [25] | 25 µg/mL |
| RNA | < 1 µg/mL [26] | 50 µg/mL |
| Glycine | <3.1 mM | 20 mM |
| HSA | 0 – 20 µg/mL | ≈ 2 µg/mL* |
| IgG | 0 – 6 µg/mL | 30 µg/mL |

**Table 1**. Selected components of urine: reference range (usually defined as the set of values 95% of the normal population falls within), and non-interfering concentrations of the same components with respect to heparin detection by Heparin Red Ultra (values taken from figure 2). "Non-interfering level" means that the fluorescence in the presence of these compounds is within ± 3.1 % of the control (compare figure 2). * Extrapolated from figure 2.

It is evident from table 1 that no or only little interference in the Heparin Red Ultra assay is expected from most of the listed urine components if present in their "normal" or reference range. A notable exception is, however, HSA which may significantly mask heparin response even at normal urinary levels (compare figure 3).

In a preliminary evaluation, we have analysed two urine samples of healthy donors, either non-spiked or spiked with 4,38 µg/mL heparin (figure 5). In one sample (urine 1), spike recovery was similar to that in water, while in the other sample (urine 2), fluorescence decrease was lower than expected and, consequently, spike recovery incomplete (figure 5). We suspected that the masking effect of HSA was the origin of incomplete spike recovery. This was confirmed by quantification of HSA in the urine samples, using an antibody-based turbidimetric detection method (Hemocue albumin 201 system). While in urine 1, HSA was not detectable (< 5 µg/mL, detection limit of the method), urine 2 displayed a HSA level of 5,5 µg/mL. The fluorescence response of the heparin–spiked urine 2 sample is in good agreement with that of an aqueous sample containing 5 µg/mL HSA (figure 3), supporting the idea that HSA partially masks the response of the assay in urine 2.

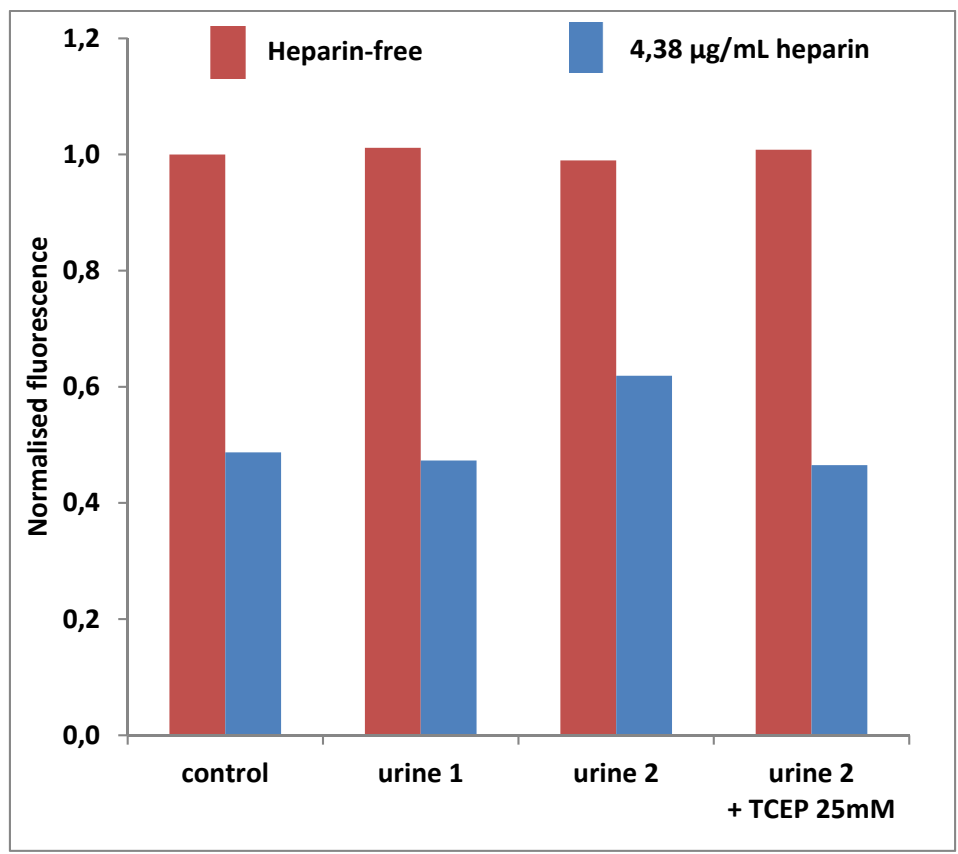

**Figure 5.** Fluorescence response of the Heparin Red Ultra assay to heparin-free urine samples 1 and 2 (red bars), and the same urine samples spiked with 4,38 µg/mL heparin (blue bars). The assay with urine 2 was repeated after pretreatment with 25 mM TCPE (incubation time 15 minutes at ambient temperature). Control (leftmost bars) refers to water or 4,38 µg/mL heparin in water, respectively. Averages of triplicate determinations; CV averaged over all concentrations: 1.3 %.

HSA was further confirmed as an interfering component in urine 2 by the effect of the disulfide-cleaving agent TCEP. Pre-treatmrent of the sample with 25 mM TCEP for 15 minutes results in a reduction of fluorescence response to heparin similar to the level observed for the control, indicating that the interference has been eliminated (compare figure 4). Additional data with a larger number of urine samples are required to confirm the reliability of the Heparin Red Ultra assay combined with TCEP sample pretreatment for heparin quantification in urine.

## Conclusion

This study addresses the need for simple and user-friendly analytical methods for the fast and accurate quantification of heparin in complex matrices. Commercially availbale Heparin Red Ultra is a ready-to-use solution that is mixed with a small volume of heparin containing sample in a microplate well for one minute, followed by fluorescence readout. A screening of typical impurities related to industrial heparin extraction from animal tissues (selected glycosaminoglycans, residual nucleic acids and proteins) as well as of components of the urine matrix (inorganic salts, amino acids, trace proteins) revealed that these compounds even in large excess have no or very little effect on the accuracy of heparin determination. Interference by human serum albumin appears to be related to extensive intramolecular disulfide bonding, a specific structural feature of this protein, and is readily overcome with a convenient sample pre-treatment protocol using the commercial disulfide cleaving agent tris(2-carboxyethyl)phosphine (TCEP). A preliminary evaluation of the fluorescence assay for heparin detection in urine, a complex multicomponent mixture, also revealed good accuracy. Heparin Red Ultra may facilitate process control of raw heparin preparation and purification in industrial manufacturing, support quality control of heparin batches, and allow convenient quantification of urinary excretion in pharmacokinetic studies.

**Conflict of interest.** R.K. holds shares in Redprobes UG, Münster, Germany. Other authors: No conflict of interest.